\documentclass[twocolumn,prd,superscriptaddress,showpacs,floatfix,%
preprintnumbers,nofootinbib]{revtex4}

\usepackage{epsfig}
\usepackage{bm}
\usepackage{multirow}

\usepackage[dvips]{color}
\definecolor{Black}{named}{Black}
\definecolor{Blue}{named}{Blue}
\definecolor{Red}{named}{Red}

\newcommand{\I}{{\rm i}}

\newcommand{\D}{{\rm d}}

\begin{document}

\title{Mu--tau neutrino refraction and collective three-flavor
transformations in supernovae}

\author{Andreu Esteban-Pretel}
\affiliation{Institut de F\'\i sica Corpuscular (CSIC--Universitat de
Val\`encia), Ed.\ Instituts d'Investigaci\'o, Ap.\ correus 22085, 46071
Val\`encia, Spain}

\author{Sergio Pastor}
\affiliation{Institut de F\'\i sica Corpuscular (CSIC--Universitat de
Val\`encia), Ed.\ Instituts d'Investigaci\'o, Ap.\ correus 22085, 46071
Val\`encia, Spain}

\author{Ricard Tom\`as}
\affiliation{Institut de F\'\i sica Corpuscular (CSIC--Universitat de
Val\`encia), Ed.\ Instituts d'Investigaci\'o, Ap.\ correus 22085, 46071
Val\`encia, Spain}

\author{Georg G.~Raffelt}
\affiliation{Max-Planck-Institut f\"ur Physik
(Werner-Heisenberg-Institut), F\"ohringer Ring 6, 80805 M\"unchen,
Germany}

\author{G{\"u}nter Sigl}
\affiliation{II.\ Institut f\"ur theoretische Physik, Universit\"at
Hamburg, Luruper Chaussee 149, 22761 Hamburg, Germany}
\affiliation{APC~\footnote{UMR 7164 (CNRS, Universit\'e Paris 7, CEA,
Observatoire de Paris)} (AstroParticules et Cosmologie), 10, rue
Alice Domon et L\'eonie Duquet, 75205 Paris Cedex 13, France}

\date{11 February 2008}

\preprint{MPP-2007-178, IFIC/07-69}

\begin{abstract}
We study three-flavor collective neutrino transformations in the
dense-neutrino region above the neutrino sphere of a supernova core.
We find that two-flavor conversions driven by the atmospheric mass
difference and the 13-mixing angle capture the full effect if one
neglects the second-order difference between the $\nu_\mu$ and
$\nu_\tau$ refractive index. Including this ``mu--tau matter term''
provides a resonance at a density of $\rho\approx3\times10^7~{\rm
g}~{\rm cm}^{-3}$ that typically causes significant modifications of
the overall $\nu_e$ and $\bar\nu_e$ survival probabilities. This
effect is surprisingly sensitive to deviations from maximal 23-mixing,
being different for each octant.
\end{abstract}


\maketitle

\section{Introduction}                        \label{sec:introduction}

Neutrinos of different flavor suffer different refraction in
matter~\cite{Wolfenstein:1977ue}. The energy shift between $\nu_e$ and
$\nu_\mu$ or $\nu_\tau$ is $\Delta V=\sqrt{2}\,G_{\rm F}Y_e n_B$ with
$G_{\rm F}$ the Fermi constant, $n_B$ the baryon density, and
$Y_e=n_e/n_B$ the electron fraction. $\Delta V$ is caused by the
charged-current $\nu_e$-electron interaction that is absent for
$\nu_\mu$ and $\nu_\tau$. For a matter density $\rho=1$ g cm$^{-3}$ we
have $\sqrt{2}\,G_{\rm F}n_B = 7.6\times10^{-14}$~eV, yet this small
energy shift is large enough to be of almost universal importance for
neutrino oscillation physics.

In normal matter, $\mu$ and $\tau$ leptons appear only as virtual
states in radiative corrections to neutral-current $\nu_\mu$ and
$\nu_\tau$ scattering, causing a shift $\Delta V_{\mu\tau}=
\sqrt{2}\,G_{\rm F}Y_\tau^{\rm eff}n_B$ between $\nu_\mu$
and~$\nu_\tau$. It has the same effect on neutrino dispersion as real
$\tau$ leptons with an abundance~\cite{Botella:1986wy}
\begin{equation}
 Y_\tau^{\rm eff}=\frac{3\sqrt{2}\,G_{\rm F}m_\tau^2}{(2\pi)^2}
 \left[\ln\left(\frac{m_W^2}{m_\tau^2}\right)-1+\frac{Y_n}{3}\right]
 =2.7\times10^{-5}\,,
\label{eq:Ytau}
\end{equation}
where $n_e=n_p$ was assumed. For the neutron abundance we have used
$Y_n=n_n/n_B=0.5$, but it provides only a 2.5\% correction so that its
exact value is irrelevant. A large nonstandard contribution to
$Y_\tau^{\rm eff}$ can arise from radiative corrections in
supersymmetric models~\cite{Roulet:1995qb}, but we will here focus on
the standard-model effect alone.

This ``mu--tau matter effect'' modifies oscillations if $\Delta
V_{\mu\tau}\agt\Delta m^2/2E$. For propagation through the Earth and
for $\Delta m^2_{\rm atm}=2$--$3\times10^{-3}~{\rm eV}^2$, this occurs
for neutrino energies $E\agt100$~TeV. The oscillation length then far
exceeds $r_{\rm Earth}$ so that $\Delta V_{\mu\tau}$ is irrelevant for
the high-energy neutrinos that are searched for by neutrino
telescopes.

\begin{figure}
\includegraphics[angle=0,width=0.9\columnwidth]{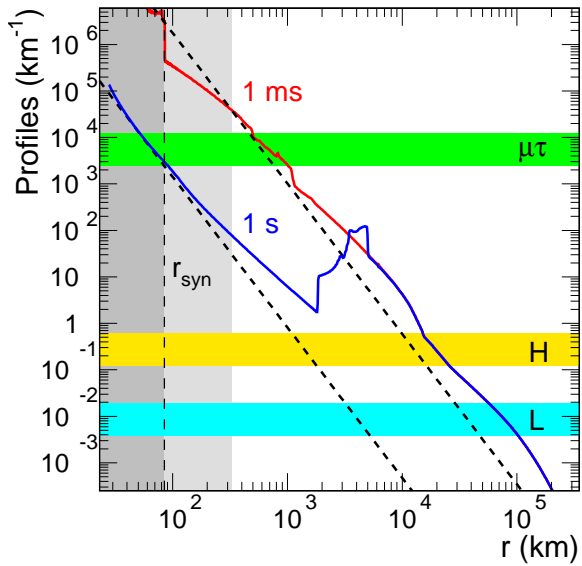}
\caption{Density profiles in terms of the weak potential $\Delta
V=\sqrt{2}\,G_{\rm F}n_e$ at 1~ms and 1~s post bounce of the numerical
SN models described in Ref.~\cite{Arcones:2006uq} (solid lines).
The dashed lines represent the simplified matter profile of
Eq.~(\ref{eq:profile}) for $\lambda_0 = 4\times 10^{6}$~km$^{-1}$ and
$\lambda_0 = 5\times 10^{9}$~km$^{-1}$, used in our numerical
calculations in Figure \ref{fig:evolution}. As horizontal bands we
indicate the conditions $\Delta V_{\mu\tau}=\Delta m^2_{\rm atm}/2E$,
$\Delta V=\Delta m^2_{\rm atm}/2E$, and $\Delta V=\Delta m^2_{\rm
sol}/2E$ for a typical range of SN neutrino energies. The gray shaded
range of radii corresponds to the region of collective neutrino
transformations. Within the radius $r_{\rm syn}$ the collective
oscillations are of the synchronized type.\label{fig:profiles}}
\end{figure}

Alternatively, the mu--tau matter effect can be important at the
large densities encountered by neutrinos streaming off a supernova
(SN) core~\cite{Akhmedov:2002zj}. For $E=20~{\rm MeV}$ the condition
$\Delta V_{\mu\tau}=\Delta m^2_{\rm atm}/2E$ implies
$\rho\approx3\times10^7~{\rm g}~{\rm cm}^{-3}$. Numerical SN density
profiles~\cite{Arcones:2006uq} reveal that this occurs far beyond the
shock-wave radius during the accretion phase, but retracts close to
the neutrino sphere after the explosion has begun. To illustrate this
point we show in Fig.~\ref{fig:profiles} the same matter density
profiles as in Ref.~\cite{Arcones:2006uq} at 1~ms post bounce (red
line) and at 1~s post bounce (blue line). As a green horizontal band
we indicate the condition $\Delta V_{\mu\tau}=\Delta m^2_{\rm
atm}/2E$ for a typical range of SN neutrino energies, whereas the
yellow and light-blue bands indicate the densities corresponding to
the H-resonance (driven by $\Delta m_{\rm atm}^2$) and the
L-resonance (driven by $\Delta m_{\rm sol}^2$). The $\nu_\mu$,
$\nu_\tau$, $\bar\nu_\mu$ and $\bar\nu_\tau$ fluxes from a SN are
virtually identical, leaving the $\mu\tau$-resonance moot, whereas
the H- and L-resonances cause well-understood consequences that are
completely described by the energy-dependent swapping probabilities
for $\nu_e$ and $\bar\nu_e$ with some combination $\nu_x$ of the
$\mu$ and $\tau$ flavor~\cite{Dighe:1999bi}. Therefore, the
traditional view has been that genuine three-flavor effects play no
role for SN neutrino oscillations unless mu and tau neutrinos are
produced with different fluxes~\cite{Akhmedov:2002zj}.

In a recent series of papers~\cite{Pastor:2002we, Sawyer:2004ai,
Sawyer:2005jk, Duan:2005cp,Duan:2006an, Hannestad:2006nj,
Duan:2007mv, Raffelt:2007yz, EstebanPretel:2007ec, Raffelt:2007cb,
Raffelt:2007xt, Duan:2007fw, Fogli:2007bk, Duan:2007bt, Duan:2007sh,
Dasgupta07} it was recognized, however, that the traditional picture
was not complete: neutrino-neutrino interactions cause large
collective flavor transformations in the SN region out to a few
100~km (gray shaded region in Fig.~\ref{fig:profiles}). With
the exception of Refs.~\cite{Duan:2007sh,Dasgupta07}, only two-flavor
conversions driven by $\Delta m^2_{\rm atm}$ and the small
$\Theta_{13}$ have thus far been studied.

We here extend our previous numerical
solutions~\cite{EstebanPretel:2007ec} to the case of three neutrino
flavors. Our main results can be summarized as follows: (i)~A
two-flavor treatment indeed captures the full effect if one
ignores~$\Delta V_{\mu\tau}$ and if the ordinary MSW resonances
occur outside of the collective neutrino region. (ii)~Including
$\Delta V_{\mu\tau}$ strongly modifies the $\nu_e$ or $\bar\nu_e$
survival probabilities, influencing the neutrino signal from the next
galactic~SN. (iii)~The effect depends sensitively on a possible
deviation from maximal~$\Theta_{23}$.  The purpose of our paper is to
provide a first illustration of these findings that no doubt need to
be refined in future.

Our work is organized as follows. In Sec.~\ref{sec:eoms} we present
the equations of motion which are solved for the three-flavor
neutrino fluxes in a simplified scenario for the SN environment. In
Sec.~\ref{sec:smallmutau} we consider the limit of a vanishing
$\mu\tau$ matter effect, while our results when it is significant are
described in Sec.~\ref{sec:largemutau}. We conclude in
Sec.~\ref{sec:conclusion}.

\section{Equations of Motion}                         \label{sec:eoms}

Mixed neutrinos are described by matrices of density $\rho_{\bf p}$
and $\bar\rho_{\bf p}$ for each (anti)neutrino mode. The diagonal
entries are the usual occupation numbers whereas the off-diagonal
terms encode phase information. The equations of motion (EOMs) are
\begin{equation}
\I\partial_t\varrho_{\bf p}=[{\sf H}_{\bf p},\varrho_{\bf p}]\,,
\label{eq:eoms}
\end{equation}
where the Hamiltonian is~\cite{Sigl:1992fn}
\begin{equation}
 {\sf H}_{\bf p}=\Omega_{\bf p}
 +{\sf V}+\sqrt{2}\,G_{\rm F}\!
 \int\!\frac{\D^3{\bf q}}{(2\pi)^3}
 \left(\varrho_{\bf q}-\bar\varrho_{\bf q}\right)
 (1-{\bf v}_{\bf q}\cdot{\bf v}_{\bf p})\,,
\label{eq:hamiltonian}
\end{equation}
${\bf v}_{\bf p}$ being the velocity. The matrix of vacuum
oscillation frequencies is $\Omega_{\bf p}={\rm
diag}(m_1^2,m_2^2,m_3^2)/2|{\bf p}|$ in the mass basis. The matter
effect is represented, in the weak interaction basis, by ${\sf
V}=\sqrt{2}\,G_{\rm F}n_B\,{\rm diag}(Y_e,0,Y_\tau^{\rm eff})$. For
antineutrinos the only difference is $\Omega_{\bf p}\to-\Omega_{\bf
p}$.

In spherical symmetry the EOMs can be expressed as a closed set of
differential equations along the radial
direction~\cite{EstebanPretel:2007ec, Fogli:2007bk}. We solve them
numerically as previously described~\cite{EstebanPretel:2007ec}, now
using $3\times3$ matrices instead of polarization vectors. The factor
$(1-{\bf v}_{\bf q}\cdot{\bf v}_{\bf p})$ in the Hamiltonian implies
``multi-angle effects'' for neutrinos moving on different
trajectories~\cite{Sawyer:2004ai, Sawyer:2005jk, Duan:2006an}.
However, for realistic SN conditions the modifications are small,
allowing for a single-angle approximation. We implement this
approximation by launching all neutrinos with $45^\circ$ relative to
the radial direction~\cite{EstebanPretel:2007ec}.

As a further simplification we use a monochromatic spectrum
($E=20~{\rm MeV}$), ignoring the ``spectral splits'' caused by
collective oscillation effects~\cite{Duan:2006an, Raffelt:2007cb,
  Raffelt:2007xt, Duan:2007fw, Fogli:2007bk}. Oscillation effects
require flavor-dependent flux differences. One expects $F_{\nu_e} >
F_{\bar\nu_e} > F_{\nu_\mu} = F_{\bar\nu_\mu} = F_{\nu_\tau} =
F_{\bar\nu_\tau}$.  The equal parts of the fluxes drop out of the
EOMs, so as initial condition we use
$F_{\nu_\mu,\bar\nu_\mu,\nu_\tau,\bar\nu_\tau}=0$ and $F_{\nu_e}
=(1+\epsilon) F_{\bar\nu_e}$ with $\epsilon=0.25$.

For the neutrino parameters we use $\Delta m^2_{12}=\Delta m^2_{\rm
  sol}=7.6\times10^{-5}~{\rm eV}^2$, $\Delta m^2_{13}=\Delta m^2_{\rm
  atm}=2.4\times10^{-3}~{\rm eV}^2$, $\sin^2\Theta_{12}=0.32$,
$\sin^2\Theta_{13}=0.01$, and a vanishing Dirac phase
$\delta=0$, all consistent with
measurements~\cite{Maltoni:2004ei,Fogli:2005cq,GonzalezGarcia:2007ib}.
We consider the entire allowed range $0.35\leq\sin^2\Theta_{23}\leq 0.65$
because our results depend sensitively on $\Theta_{23}$.

We use a fixed matter profile of the form $\rho\propto r^{-3}$,
implying a radial variation of the weak potential~of
\begin{equation}
\Delta V=Y_e\lambda_0\,\left(\frac{R}{r}\right)^3\,,
\label{eq:profile}
\end{equation}
where $R=10$~km is our nominal neutrino-sphere radius and
$Y_e=0.5$. In Fig.~\ref{fig:profiles} we show this profile (dashed
lines) for two different values of $\lambda_0= 4\times
10^{6}$~km$^{-1}$ and $\lambda_0 = 5\times 10^{9}$~km$^{-1}$. For the
former case, the H-resonance is at $r_{\rm H}=1.9\times 10^3 $~km, the
L-resonance at $r_{\rm L}=8.3\times 10^3$~km, and the
$\mu\tau$-resonance at $r_{\mu\tau}=71$~km. For the latter they are at
$r_{\rm H}=2.0\times 10^4 $~km, $r_{\rm L}=9.0\times 10^4$~km, and
$r_{\mu\tau}=760$~km.\footnote{We loosely refer to the radius where
$\Delta m_{\rm atm}^2/2E=\Delta V_{\mu\tau}$ as the $\mu\tau$
resonance, although this would be correct only for a small vacuum
mixing angle in the 23-subsystem.}

The strength of the neutrino-neutrino interaction can be
parametrized by
\begin{equation}
\mu_0 = \sqrt{2}G_{\rm F}(F^R_{\bar\nu_e}-F^R_{\bar\nu_x})\,,
\end{equation}
where the fluxes are taken at the neutrino-sphere radius~$R$. As in
our previous work~\cite{EstebanPretel:2007ec} we shall assume
$\mu_0=7\times 10^5$~km$^{-1}$. In the single-angle approximation
where all neutrinos are launched with $45^\circ$ relative to the
radial direction~\cite{EstebanPretel:2007ec}, the radial dependence
of the neutrino-neutrino interaction strength can be explicitly
written as
\begin{equation}
\mu(r) = \mu_0 \frac{R^4}{r^4}\frac{1}{2-R^2/r^2}\,.
\end{equation}
 While the $r^{-4}$ scaling of $\mu(r)$ for $r\gg R$ is
generic, the overall strength $\mu_0$ depends on the neutrino fluxes
and on their angular divergence, i.e., on the true radius of the
neutrino sphere. Our $R=10$~km is not meant to represent the physical
neutrino sphere, it is only a nominal radius where we fix the inner
boundary condition for our calculation.

The collective neutrino oscillations are of the synchronized
type within the ``synchronization radius.'' For our chosen $\mu_0$
and for the assumed excess $\nu_e$ flux of 25\% we find $r_{\rm
syn}\simeq 100$~km as indicated in Fig.~\ref{fig:profiles}. Collective
flavor transformations occur at $r>r_{\rm syn}$. Therefore, the
$\mu\tau$ matter effect can be important only if it is sufficiently
large for $r>r_{\rm syn}$.

Figure~\ref{fig:profiles} illustrates that the region where the
$\mu\tau$-resonance takes place depends on the time after bounce. For
realistic values of the matter density profile and neutrino-neutrino
interaction, one expects $r_{\mu\tau}$ to lie far beyond the
collective region at early times. This can be inferred from the
relative position of $r_{\rm syn}$ and the intersection of the 1~ms
profile and the green band. At later times though the proto neutron
star contracts and $r_{\mu\tau}$ moves to smaller radii. Eventually
$r_{\mu\tau}$ becomes smaller than $r_{\rm syn}$, at which point
$\Delta V_{\mu\tau}$ becomes irrelevant.

In order to mimic these different situations we will
us a simple power-law matter profile of the form in
Eq.~(\ref{eq:profile}). In other words, we will use a mu-tau matter
potential of the form
\begin{equation}
\Delta V_{\mu\tau}=Y_{\tau}^{\rm eff}\lambda_0\,\left(\frac{R}{r}\right)^3\,,
\label{eq:mutaueff}
\end{equation}
with a fixed $Y_{\tau}^{\rm eff}$ given by Eq.~(\ref{eq:Ytau}) and a
variable coefficient $\lambda_0$. Therefore early and late times can
be reproduced by considering large and small values of $\lambda_0$,
respectively, as can be seen in Fig.~\ref{fig:profiles}. In other
words, we will always assume that the ordinary MSW resonances are far
outside of the collective neutrino region, whereas the $\mu\tau$
resonance can lie at smaller (vanishing $\mu\tau$ matter effect) or
larger (large $\mu\tau$ matter effect) radii than~$r_{\rm syn}$.

\section{Vanishing mu-tau matter effect}        \label{sec:smallmutau}

\begin{figure*}[ht]
\includegraphics[angle=0,width=1.0\textwidth]{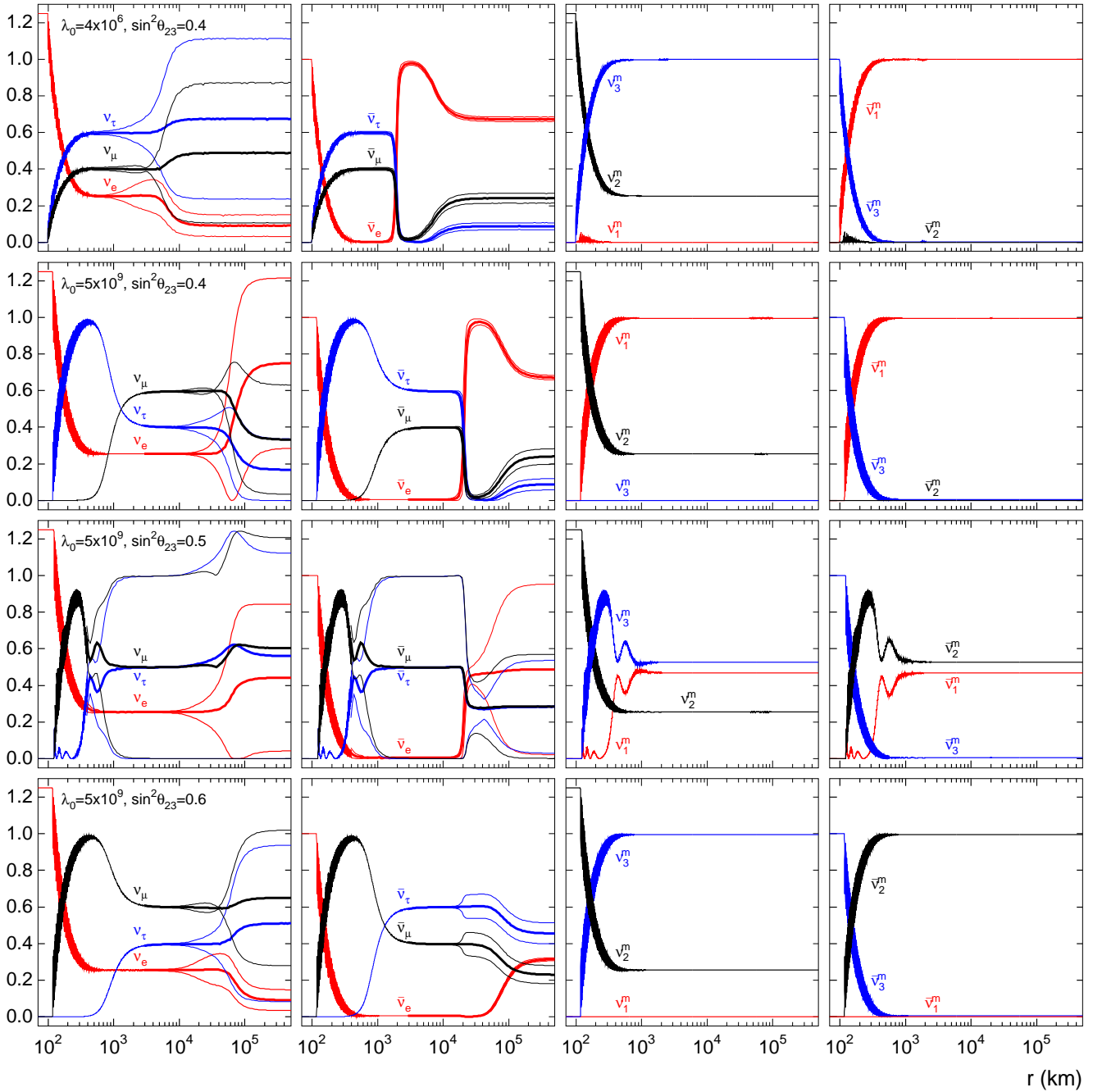}
\caption{Radial evolution of the neutrino fluxes, normalized to the
initial $\bar{\nu}_e$ flux, for a fixed neutrino energy ($E_\nu=20$
MeV) and an inverted $\Delta m^2_{\rm atm}$. {}From left to right:
neutrino weak eigenstates, antineutrino weak eigenstates, neutrino
propagation eigenstates and antineutrino propagation eigenstates. In
the first two columns, after bipolar conversions we
show the average as thick lines and the envelopes of the
fast-oscillating curves as thin lines. The top row shows the case of a
vanishing $\mu\tau$ matter effect, while the three bottom rows use a
large $\mu\tau$ effect with different values for the 23 mixing angle
as indicated.\label{fig:evolution}}
\end{figure*}

As a first case we consider the traditional assumption of a vanishing
$\mu\tau$ matter effect, which we account for using a value of
$\lambda_0=4\times 10^6$. We assume an inverted $\Delta m^2_{\rm atm}$
and use a non-maximal value $\sin^2\Theta_{23}=0.4$.  Our numerical
calculations for this case are shown in the top row of
Fig.~\ref{fig:evolution}. The first two panels correspond to the
radial evolution of the fluxes of the weak interaction eigenstates of
neutrinos and antineutrinos, respectively, whereas in the last two
panels we show the evolution of the propagation eigenstates.  These
are the eigenstates of $\Omega_{\bf p} +{\sf V}$, i.e., of that part
of the Hamiltonian Eq.~(\ref{eq:hamiltonian}) that does not include
the neutrino-neutrino interactions. In the collective neutrino region,
we observe the usual pair conversion of the $\nu_e$ and $\bar\nu_e$
fluxes into the $\mu$ and $\tau$ flavors. Had we chosen a maximal $23$
mixing angle, the appearance curves for these flavors would be
identical.

For larger distances the evolution consists of ordinary MSW
transformations that are best pictured in the basis of instantaneous
propagation eigenstates in matter (last two
panels). Beyond the collective transformation region, all neutrinos
and antineutrinos stay fixed in their propagation eigenstates. In the
weak-interaction basis, on the other hand, this implies fast
oscillations because we have a fixed energy, preventing kinematical
decoherence between different energy modes. In the panels for neutrino
and antineutrino interaction states, for radii beyond the
dense-neutrino region we show as thick lines the average evolution as
well as the envelopes of the fast-oscillating flavor fluxes.

\begin{figure}[ht]
\includegraphics[angle=0,width=0.95\columnwidth]{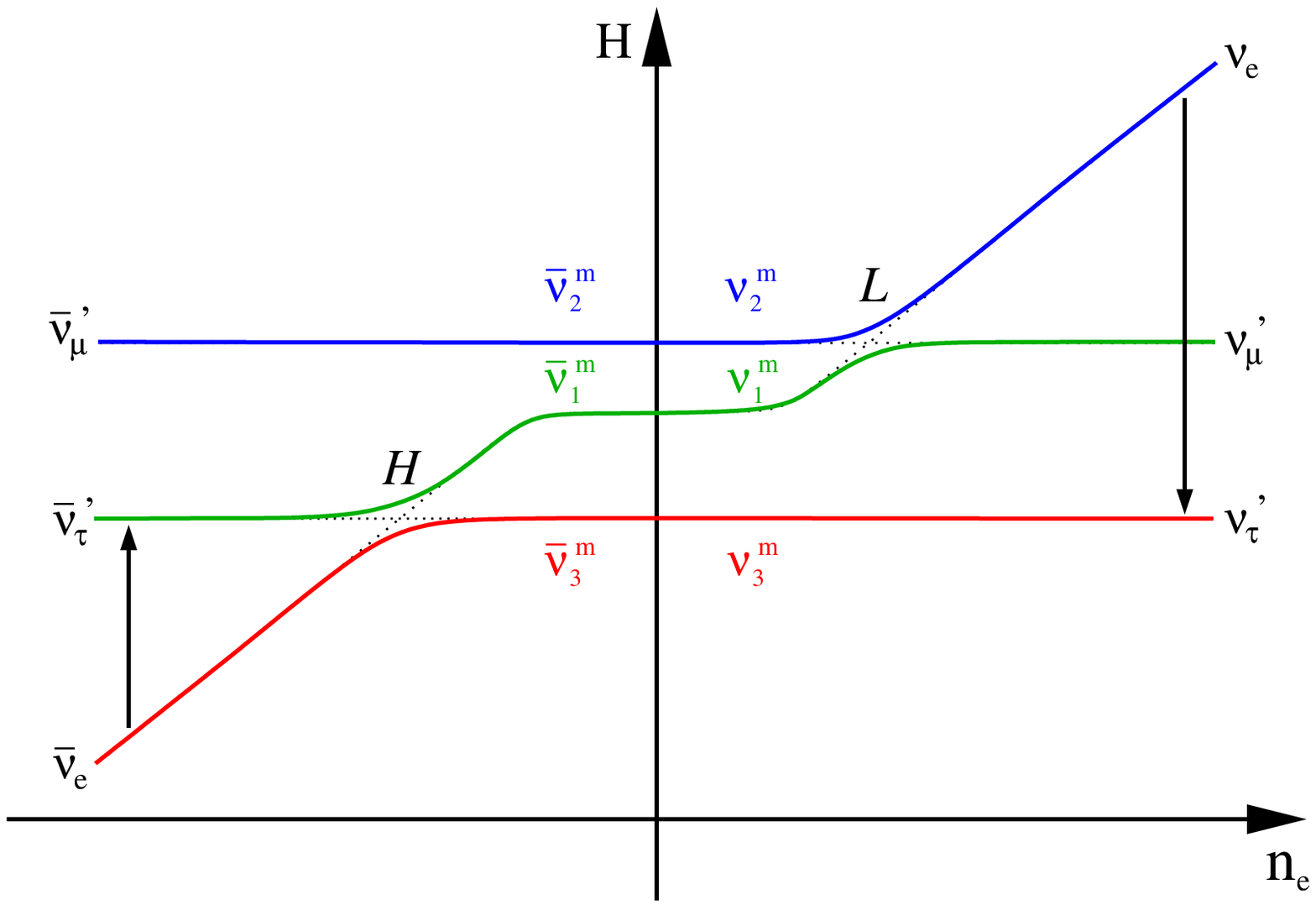}
\vskip24pt
\includegraphics[angle=0,width=0.95\columnwidth]{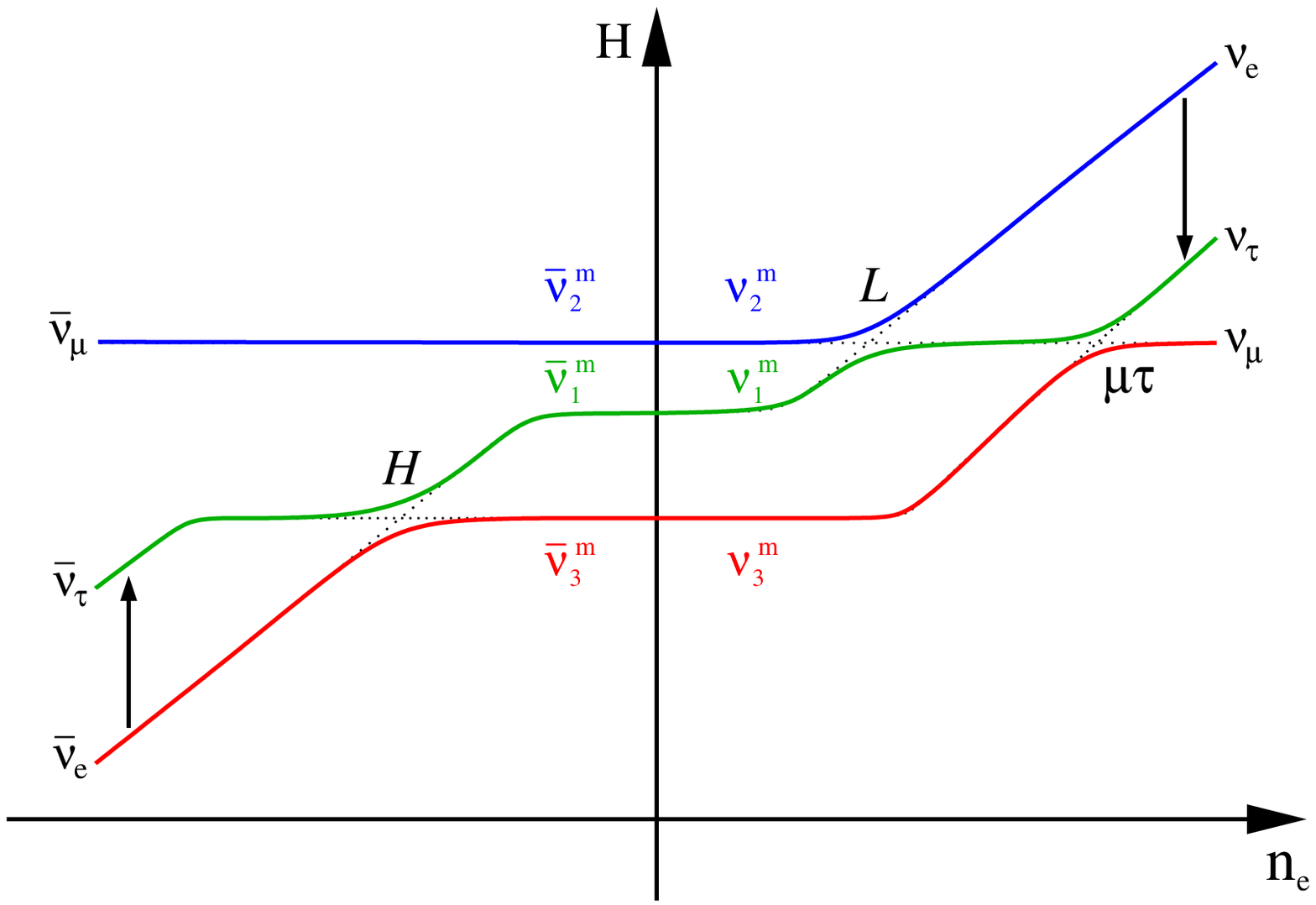}
\vskip24pt
\includegraphics[angle=0,width=0.95\columnwidth]{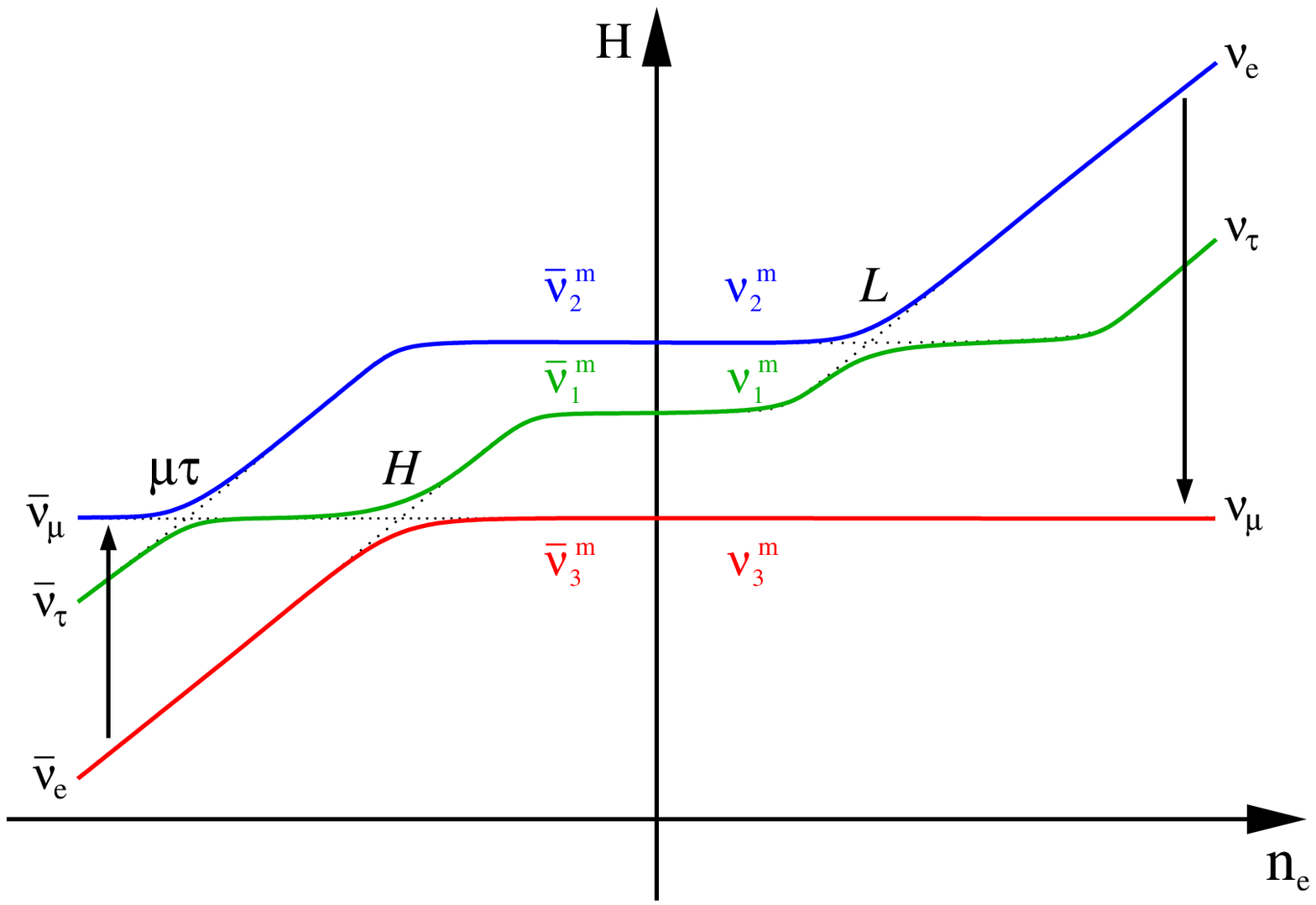}
\caption{Level crossing scheme of neutrino conversion for the inverted
hierarchy in a medium with a vanishing $\Delta V_{\mu\tau}$ (upper
panel) and a large $\Delta V_{\mu\tau}$ with 23-mixing in the first
octant (central panel) or the second octant (lower panel).  The arrows
indicate the transitions caused by collective flavor transformations.
\label{fig:crossing}}
\end{figure}

Another way of describing this evolution is by the level crossing
schemes of Fig.~\ref{fig:crossing}. The upper panel represents the
case of vanishing $\Delta V_{\mu\tau}$, corresponding to Fig.~5d of
Ref.~\cite{Dighe:1999bi}. The central panel represents the case with
large $\Delta V_{\mu\tau}$ and a 23-mixing angle in the first octant
and is similar to Fig.~2 of Ref.~\cite{Akhmedov:2002zj}.  In such
plots one shows the neutrino energy levels as a function of the matter
density. The continuation of this diagram to negative densities gives
us the energy levels of antineutrinos: the neutrino energy at a
negative density really means the antineutrino energy at the
corresponding positive density. For vanishing density (vacuum), we
have the three vacuum mass eigenstates that are identical for
neutrinos and antineutrinos.  The upper (blue) line corresponds to
propagation eigenstate~2, the middle (green) line to~1, and the bottom
(red) line to~3, a scheme representing the inverted hierarchy
case. These lines represent the propagation eigenstates that are
adiabatically connected for different densities.

While in vacuum the propagation eigenstates coincide with the mass
eigenstates, at large densities they correspond to weak interaction
eigenstates. For vanishing $\Delta V_{\mu\tau}$ and at the low
energies relevant to our problem, the $\mu$ and $\tau$ flavor are not
distinguishable so that any convenient linear combination can be
chosen as interaction eigenstates. It is convenient to introduce the
states $\nu_\mu'$ and $\nu_\tau'$ that correspond to a vanishing
23-mixing angle, i.e., they diagonalize the 23-subsystem. If the small
13-mixing angle were to vanish, the 3-mass eigenstate would coincide
with $\nu_\tau'$. In the upper panel of Fig.~\ref{fig:crossing} and
using the $(\nu_e,\nu_\mu',\nu_\tau')$ basis, the 2-state connects
adiabatically to $\nu_e$ and $\bar\nu_\mu'$, whereas the 3-state
connects adiabatically to $\bar\nu_e$ and $\nu_\tau'$.

At the neutrino sphere, the fluxes are prepared in $\nu_e$ and
$\bar\nu_e$ eigenstates, which in the case of inverted mass hierarchy
coincide with the propagation (or matter) eigenstates $\nu^{\rm m}_2$
and $\bar\nu^{\rm m}_3$, respectively.  In the absence of
neutrino-neutrino interactions, since the L-resonance is
always adiabatic, the $\nu_e$'s leave the star as $\nu_2$. In the case
of $\bar\nu_e$ the evolution depends on $\sin^2\Theta_{13}$
\cite{Dighe:1999bi}. For values larger than $10^{-3}$ they propagate
also adiabatically (MSW transformation) and escape as $\bar\nu_3$,
whereas for values smaller than $10^{-5}$ the transition at the
H-resonance is strongly non-adiabatic: there is a jump of matter
eigenstates from $\bar\nu^{\rm m}_3$ to $\bar\nu^{\rm m}_1$ and
the $\bar\nu_e$'s leave the star as $\bar\nu_1$. As a
consequence, the survival probability is $P(\nu_e\rightarrow \nu_e)
\approx \sin^2\Theta_{12}$ and $P(\bar\nu_e\rightarrow \bar\nu_e)
\approx \sin^2\Theta_{13}$ or $\cos^2\Theta_{12}$ for large and small
$\Theta_{13}$, respectively.

In the presence of neutrino-neutrino interactions, important
collective effects take place in the inner SN layers, where the
neutrino density is high.  We observe in the first two panels of
Fig.~\ref{fig:evolution} that collective pair transformations convert
the $\nu_e$ and $\bar\nu_e$ fluxes to $\nu_\tau'$ and $\bar\nu_\tau'$
as indicated by the arrows in the upper panel of
Fig.~\ref{fig:crossing}.  The consequences for the subsequent
evolution are dramatic. In the case of $\nu_e$ a fraction equal to
$\epsilon F_{\bar\nu_e}$ stays in $\nu_2^{\rm m}$ and evolves as in
the absence of neutrino-neutrino interactions, while the rest of
$\nu_e$ are transformed to $\nu_3^{\rm m}$. As a consequence, the
final $\nu_e$ flux, normalized to the initial $\bar\nu_e$ one, is
expected to be approximately $\epsilon \sin^2\Theta_{12}\simeq 0.08$,
see thick line in the upper left panel in Fig.~\ref{fig:evolution}.
In the case of antineutrinos the effect of the collective pair
conversion is to interchange the eigenstates in which $\bar\nu_e$ and
$\bar\nu_\tau'$ arrive at the H-resonance. Now $\bar\nu_e$ enters the
resonance as $\bar\nu_1^{\rm m}$ instead of $\bar\nu_3^{\rm
  m}$. Therefore, for $\sin^2\Theta_{13}\gtrsim 10^{-3}$ the resonance
is adiabatic and the $\bar\nu_e$'s leave the star as $\bar\nu_1$,
leading to a final normalized flux of approximately
$\cos^2\Theta_{12}\simeq 0.68$, see the thick line in the second panel
in Fig.~\ref{fig:evolution}. Instead, if $\sin^2\Theta_{13}\lesssim
10^{-5}$ again there is a jump of matter eigenstates from
$\bar\nu^{\rm m}_1$ to $\bar\nu^{\rm m}_3$ at the H-resonance.  In
this case $\bar\nu_e$ leaves the star as $\bar\nu_3$, leading to a
normalized $\bar\nu_e$ flux equal to $\sin^2\Theta_{13}$.

\begin{figure}
\includegraphics[angle=0,width=0.6\columnwidth]{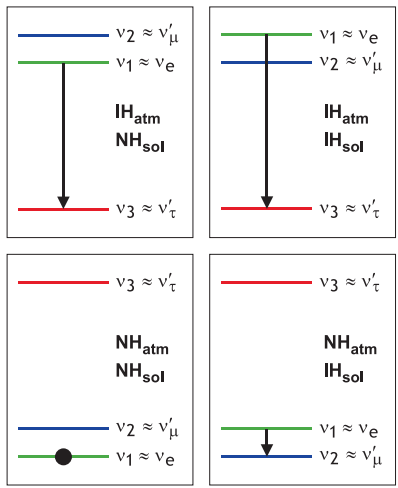}
\caption{Vacuum level diagram for all hypothetical combinations of
atmospheric and solar mass hierarchies (normal or inverted). The 12
and 13 mixing angles are assumed to be very small, mimicking the
effect of ordinary matter. The effect of collective conversions is
indicated by an arrow.\label{fig:levels}}
\end{figure}

The impact of collective effects is easier to understand if we follow
the previous literature~\cite{Duan:2005cp, Hannestad:2006nj} and
observe that, in a two-flavor system, the impact of ordinary matter
can be transformed away by going into a rotating reference frame
for the polarization vectors. Collective conversions proceed in the
same way as they would in vacuum, except that the effective mixing
angle is reduced. Therefore, assuming an inverted hierarchy (IH) for
the atmospheric mass splitting and a normal hierarchy (NH) for the
solar splitting, we should consider the level scheme as in the upper
left panel of Fig.~\ref{fig:levels}. The mass eigenstates now
approximately coincide with the interaction eigenstates because the
23-mixing angle was removed by going to the primed states, and the
mixing angles involving $\nu_e$ are effectively made small by the
presence of matter. Of course, this level scheme does not
adiabatically connect to the true vacuum situation.

The initial state consists of $\nu_e$ and $\bar\nu_e$ and thus
essentially of $\nu_1$ and $\bar\nu_1$. Collective conversions
driven by $\Delta m^2_{\rm atm}$ then transform $\nu_1\bar\nu_1$
pairs to $\nu_3\bar\nu_3$ pairs in the familiar two-flavor way. If
both hierarchies are normal, we begin in the lowest-lying state and
nothing happens.
In the hypothetical case where both hierarchies are inverted (upper
right panel in Fig.~\ref{fig:levels}), we begin in the highest state
and $\Delta m^2_{\rm atm}$ drives us directly to the lowest state.
Finally, if the atmospheric hierarchy is normal and the solar one is
inverted (lower right panel in Fig.~\ref{fig:levels}), collective
transformations driven by $\Delta m^2_{\rm sol}$ take us to the
lowest state.

We have numerically solved the evolution of the three-flavor system
with a realistic SN matter profile and found that the results confirm
this simple picture. In a two-flavor treatment, the much smaller
$\Delta m^2_{\rm sol}$ leads to collective transformations at a much
larger radius than $\Delta m^2_{\rm atm}$. In a three-flavor
treatment, $\Delta m^2_{\rm atm}$ therefore acts first and takes us
directly to the lowest-lying state if the atmospheric hierarchy is
inverted. Otherwise only the hypothetical case of the lower-right
panel in Fig.~\ref{fig:levels} is an example where $\Delta m^2_{\rm
sol}$ plays any role. We have numerically verified that normal
$\Delta m_{\rm atm}^2$ combined with inverted $\Delta m_{\rm sol}^2$
is the only case where $\Delta m_{\rm sol}^2$ drives collective
transformations. Since $\Delta m_{\rm sol}^2$ is measured to be
normal, the previous two-flavor treatments based on $\Delta m_{\rm
atm}^2$ and $\Theta_{13}$ fortuitously capture the full effect.

We conclude that in the limit of a vanishing $\mu\tau$ effect the
collective flavor transformations and the subsequent MSW evolution
factorize and that the collective effects are correctly treated in a
two-flavor picture. Of course, this situation may change if the matter
profile is so shallow that the ordinary MSW effects occur in the same
region as the collective phenomena~\cite{Duan:2007sh}.

\section{Large mu-tau matter effect}            \label{sec:largemutau}

\begin{figure}[t]
\includegraphics[angle=0,width=1.0\columnwidth]{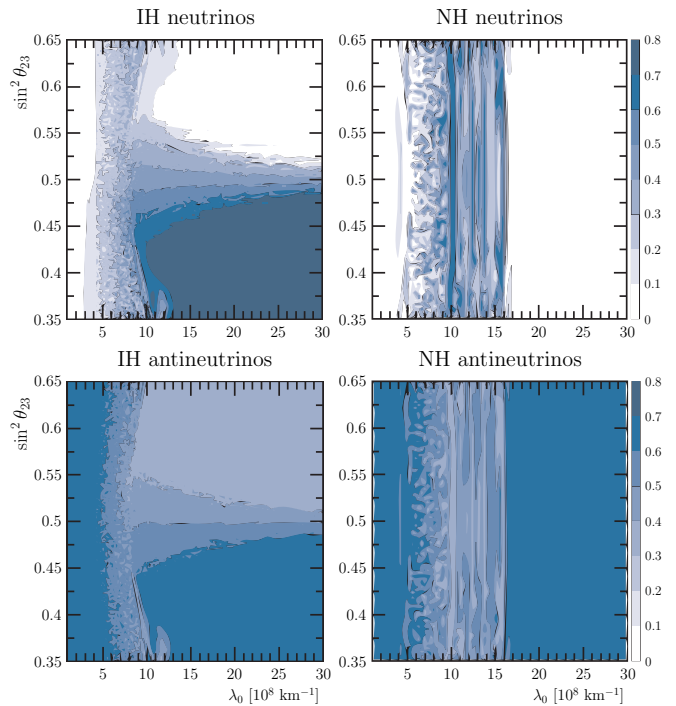}
\caption{Contours in the space of $\sin^2\Theta_{23}$ and $\lambda_0$
  for the $\nu_e$ (top) and $\bar\nu_e$ (bottom) fluxes emerging from
  the SN surface for both normal (right) and inverted (left) mass
  hierarchy. All fluxes are normalized to the initial $\bar{\nu}_e$
  flux.  We show values averaged over fast vacuum
  oscillations. \label{fig:contours}}
\end{figure}

Next we calculate the flavor evolution for the same model, now
including a significant $\Delta V_{\mu\tau}$, i.e. we assume a large
$\lambda_0$. In this case the flavor content of the neutrino and
antineutrino fluxes emerging from the SN surface depend on the
strength of $\Delta V_{\mu\tau}$ as well as the choice of
$\Theta_{23}$, as can be seen in the corresponding panels of
Fig.~\ref{fig:evolution}. This dependence is best illustrated with the
help of the contour plot Fig.~\ref{fig:contours} where we show the
$\nu_e$ and $\bar\nu_e$ fluxes emerging from the SN, averaged over
fast vacuum oscillations.

If $\Delta V_{\mu\tau}$ is so large that the mu-tau effect is strong
in the region of collective neutrino oscillations, there are two
stable limiting cases, depending on the 23 mixing angle. If the
mixing angle is sufficiently non-maximal and in the first octant, the
collective oscillations transform the initially prepared $\nu_e$ and
$\bar\nu_e$ fluxes to the propagation eigenstates as indicated by the
arrows in the middle panel of Fig.~\ref{fig:crossing}, i.e., we
observe pair transformations to $\nu_\tau\bar\nu_\tau$.

This behavior is understood if we assume that in the $\mu\tau$ system
we can once more go to a rotating frame and now simply imagine that
the 23 mixing angle is effectively small by the impact of the
$\mu\tau$ matter effect. In this case $\nu_3\approx \nu_\tau$. Since
collective quasi-vacuum oscillations take us to the lowest-lying
state, the $\nu_3$ state in the inverted hierarchy, we are
effectively taken to $\nu_\tau\bar\nu_\tau$ pairs. Instead, if the 23
mixing angle is in the second octant, $\nu_\mu$ and $\nu_\tau$ switch
roles, explaining that now $\nu_3\approx\nu_\mu$ and
$\bar\nu_3\approx\bar\nu_\mu$.

These are only heuristic explanations. We expect that they can
  be made precise in a true analytic three-flavor treatment of
  collective neutrino oscillations along the lines of
  Ref.~\cite{Dasgupta07}.

\begin{figure}
\includegraphics[angle=0,width=0.8\columnwidth]{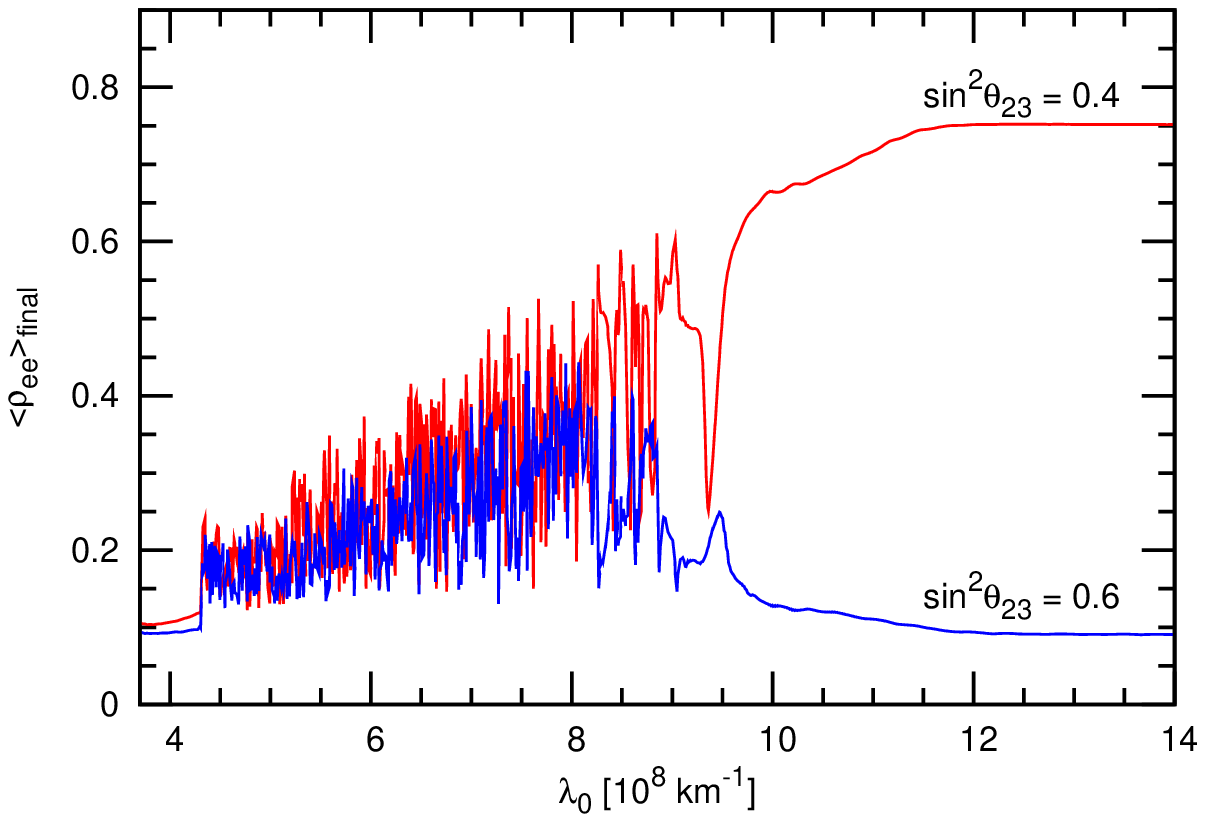}
\vskip12pt
\includegraphics[angle=0,width=0.8\columnwidth]{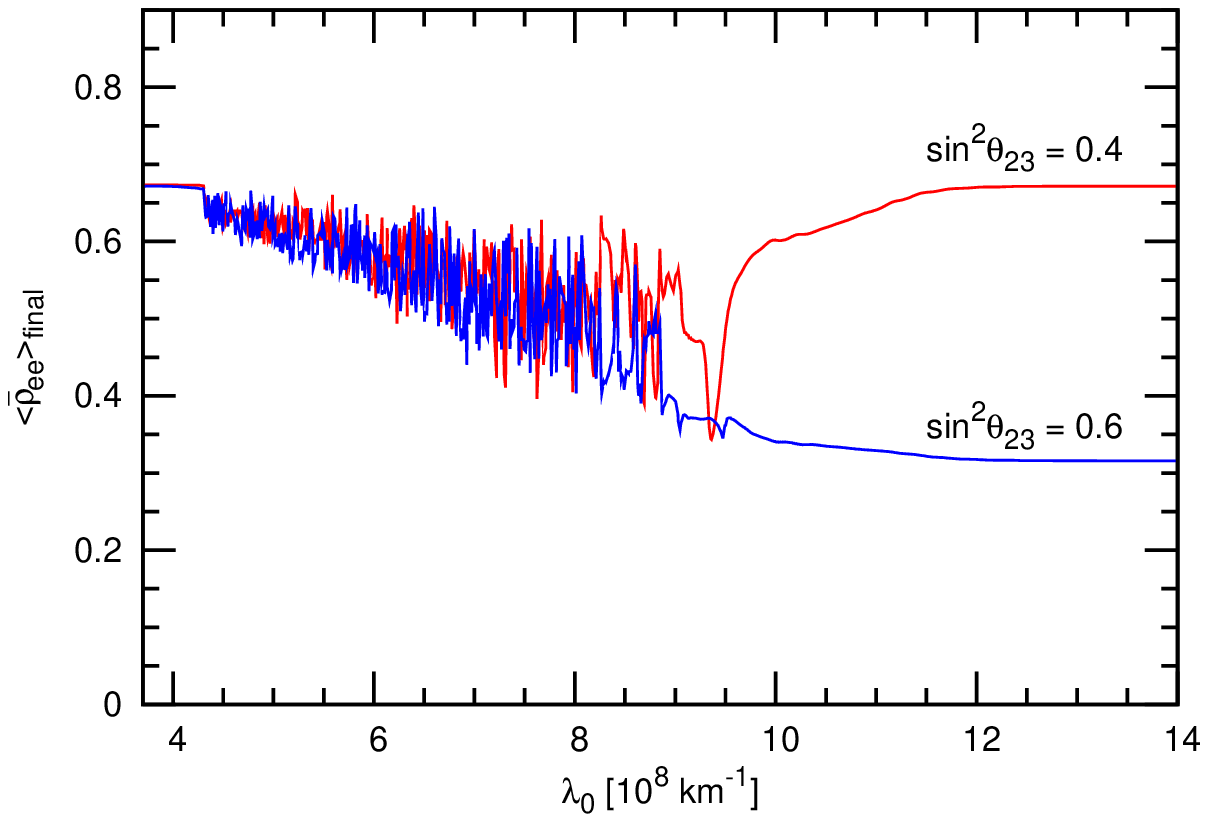}
\caption{Fluxes of $\nu_e$ (top) and $\bar\nu_e$ (bottom), normalized
  to the initial $\bar{\nu}_e$ flux, emerging from the SN as a
  function of $\lambda_0$ for a 23 mixing angle in
  the first (red line) or second (blue line) octant. These curves
  represent cuts through the inverted hierarchy contour plots of
  Fig.~\ref{fig:contours} at the indicated values of
  $\sin^2\Theta_{23}$.\label{fig:cuts}}
\end{figure}

\begin{figure}
\includegraphics[angle=0,width=0.8\columnwidth]{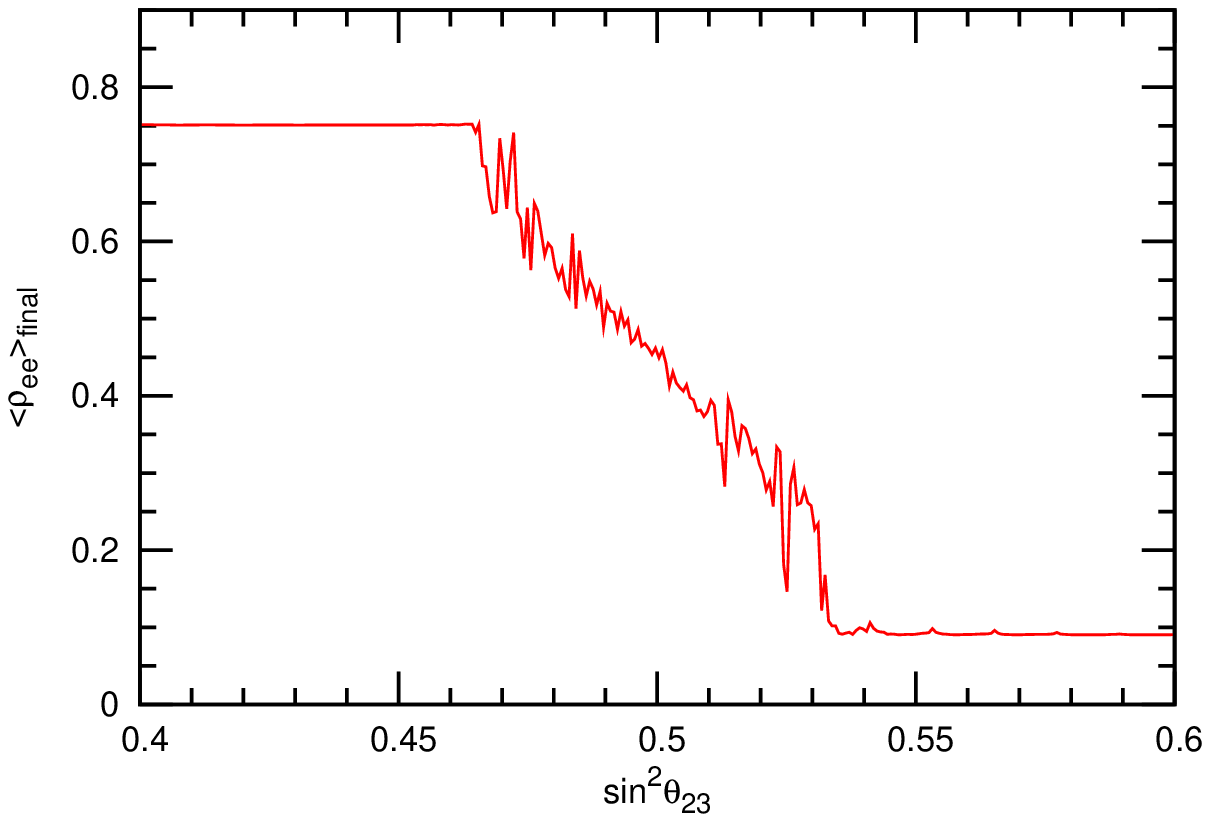}
\vskip12pt
\includegraphics[angle=0,width=0.8\columnwidth]{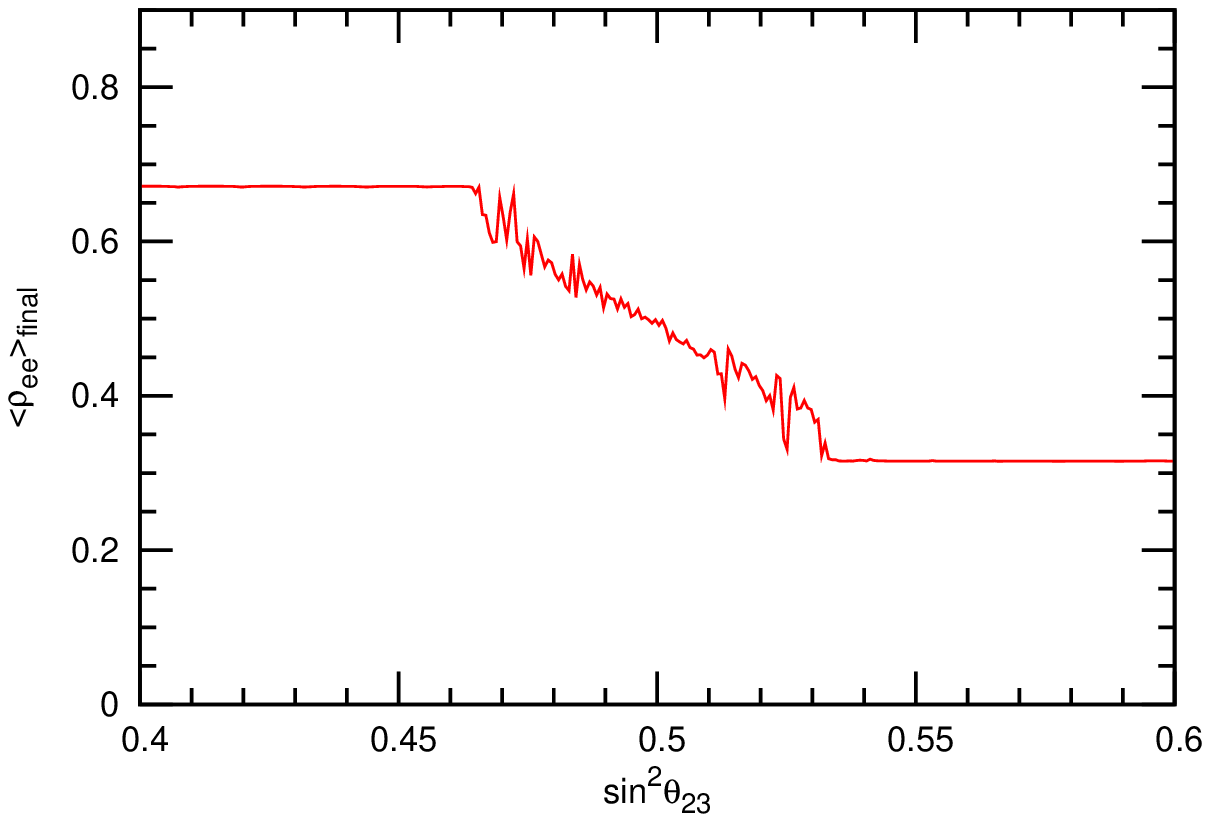}
\caption{Fluxes of $\nu_e$ (top) and $\bar\nu_e$ (bottom), normalized
to the initial $\bar{\nu}_e$ flux, emerging from the SN as a function
of $\sin^2\Theta_{23}$ for $\lambda_0= 1.85\times
10^9$~km$^{-1}$. These curves represent cuts through the inverted
hierarchy contour plots of Fig.~\ref{fig:contours} at the indicated
value of $\lambda_0$.\label{fig:cuts2}}
\end{figure}

For intermediate values of $\Delta V_{\mu\tau}$ and for 23 mixing
angles near maximal, the final fluxes depend sensitively on
parameters. For intermediate values of $\Delta V_{\mu\tau}$, there are
also nontrivial effects for the normal hierarchy. The collective
effects do not place the ensemble into propagation eigenstates,
preventing a simple interpretation. The sensitive dependence for
intermediate $\Delta V_{\mu\tau}$ is also illustrated in
Fig.~\ref{fig:cuts} where we show the emerging average $\nu_e$ and
$\bar\nu_e$ fluxes as functions of $\lambda_0$ for two values of
$\Theta_{23}$, one in the first and the other in the second octant. In
Fig.~\ref{fig:cuts2} we show the same $\nu_e$ and $\bar\nu_e$ fluxes
as functions of $\sin^2\Theta_{23}$ for $\lambda_0= 1.85\times
10^9$~km$^{-1}$. One can notice how the fall of $\bar\rho_{ee}$ is not
exactly centered at $\sin^2\theta_{23}=0.5$ but slightly shifted to
smaller values. This is due to second-order corrections to the
$\mu\tau$ resonance condition.

This dependence on the $\Theta_{23}$ octant leads to a clear imprint
on the final survival probability.  Let us first consider the first
octant. In the case of $\nu_e$ a fraction equal to $\epsilon
F_{\bar\nu_e}$ stays in $\nu_2^{\rm m}$.  However the presence of the
$\mu\tau$-resonance in the neutrino channel makes the rest of the
$\nu_e$ to be transformed to $\nu_1^{\rm m}$. 
Their subsequent evolution would depend on the
adiabaticity of the
$\mu\tau$-resonance, but it has been shown to be always
adiabatic~\cite{Akhmedov:2002zj}. As a consequence, the 
final $\nu_e$ flux is expected to be approximately
$\cos^2\Theta_{12}+\epsilon \sin^2\Theta_{12} \simeq 0.76$, see thick
line in the left panel of the second row in
Fig.~\ref{fig:evolution}. In the case of antineutrinos the situation
is completely analogous to the case of vanishing $\Delta V_{\mu\tau}$
so that $P(\bar\nu_e\rightarrow \bar\nu_e) \approx \cos^2\Theta_{12}$
or $\sin^2\Theta_{13}$, depending on the value of $\Theta_{13}$.

If $\Theta_{23}$ belongs to the second octant, then the
$\mu\tau$-resonance lies in the antineutrino channel. The crucial
point is that now all $\bar\nu_e$ are transformed to
$\bar\nu_\mu=\bar\nu_2^{\rm m}$ before reaching the
$\mu\tau$-resonance, see the lower panel in Fig.~\ref{fig:crossing}.
Taking into account that $\bar\nu_2^{\rm m}$ does not encounter the
H-resonance, the survival probability will be always
$P(\bar\nu_e\rightarrow \bar\nu_e) \approx \sin^2\Theta_{12}$,
independently of the value of $\Theta_{13}$. On the other hand
neutrinos do not feel the $\mu\tau$-resonance and therefore their
propagation is the same as in the vanishing $\Delta V_{\mu\tau}$ case.

We present in Table~\ref{tab:barnueP} a summary of the cases discussed
so far. One can see the importance of the presence of collective
neutrino effects, as well as the dependence on the strength of the
mu-tau matter effect.

\begin{table}
\caption{Summary of the approximate values of the $\bar\nu_e$ survival
probability for an inverted hierarchy, including or not collective
effects. Here a small (large) mixing angle $\Theta_{13}$
stands for $\sin^2\Theta_{13}\lesssim 10^{-5}$
($\sin^2\Theta_{13}\gtrsim 10^{-3}$), while a small (large) $\Delta
V_{\mu\tau}$ represents $r_{\mu\tau}$ being smaller
(larger) than~$r_{\rm syn}$.
\label{tab:barnueP}}
\begin{ruledtabular}
\begin{tabular}{cccccc}
Collective & \multirow{2}{*}{$\Delta V_{\mu\tau}$} &
\multirow{2}{*}{$\Theta_{23}$} &
\multirow{2}{*}{$\Theta_{13}$} &
$\bar\nu_e$ & \multirow{2}{*}{$P(\bar\nu_e\rightarrow \bar\nu_e)$}\\
effects & & & & leaves as & \\
\hline
no & any & any & small & $\bar\nu_1$ & $\cos^2\Theta_{12}$\\
no & any & any & large & $\bar\nu_3$ & $\sin^2\Theta_{13}$\\
yes & small & any & small & $\bar\nu_3$ & $\sin^2\Theta_{13}$\\
yes & small  & any & large & $\bar\nu_1$ & $\cos^2\Theta_{12}$\\
yes & large & $<\pi/4$ & small & $\bar\nu_3$ & $\sin^2\Theta_{13}$\\
yes & large  & $<\pi/4$ & large & $\bar\nu_1$ & $\cos^2\Theta_{12}$\\
yes & large  & $>\pi/4$ & any & $\bar\nu_2$ & $\sin^2\Theta_{12}$\\
\end{tabular}
\end{ruledtabular}
\end{table}

Another interesting feature concerns the position of
$r_{\rm syn}$ in the presence of a large $\mu\tau$ matter effect.  As
can be seen comparing the first two rows of Fig.~\ref{fig:evolution},
the radius where collective neutrino transformations begin is slightly
larger ($r_{\rm syn}\simeq115$~km) when we include a
significant $\Delta V_{\mu\tau}$. We have checked that, while for a
small $\Delta V_{\mu\tau}$ the position of $r_{\rm syn}$ is
independent of $\Theta_{23}$, for a large $\Delta V_{\mu\tau}$ the
onset of bipolar transformations is delayed for nonzero values of
$\Theta_{23}$. This effect is largest for maximal mixing
($\Theta_{23}=\pi/4$) and symmetric relative to $\Theta_{23}=\pi/4$.
A full understanding of this variation presumably requires an
analytic three-flavor treatment in the spirit of
Ref.~\cite{Dasgupta07}.

\section{Conclusions}                           \label{sec:conclusion}

At the relatively low energies relevant for SN neutrinos, charged mu
and tau leptons cannot be produced so that mu- and tau-flavored
neutrinos are not distinguishable in the SN or in detectors. (In the
inner core of a SN the temperatures may be high enough to produce a
significant thermal muon density, but this would not affect the
emission from the neutrino sphere.) The impact of the small
second-order difference between the $\nu_\mu$ and $\nu_\tau$
refractive index does not produce observable effects as long as one
only considers the traditional MSW flavor
conversion~\cite{Akhmedov:2002zj}.

The picture changes if one includes the unavoidable effect of
collective neutrino transformations in the region above the neutrino
sphere. If the matter density is large enough that $\Delta
V_{\mu\tau}$ is comparable to or larger than $\Delta m_{\rm
atm}^2/2E$, the survival probability of $\nu_e$ and $\bar\nu_e$ can be
completely modified and depends sensitively on the mixing angle
$\Theta_{23}$.  In future one should also include non-monochromatic
energy spectra, leading to spectral split phenomena that could be more
complicated than the previously studied two-flavor cases. One should
also explore the impact of realistic angular distributions and of a
non-zero Dirac phase in the neutrino mixing matrix.

Lower-mass progenitors may collapse with a O-Ne-Mg core and, on the
computer, explode easily because there is very little mass in the
envelope~\cite{Kitaura:2005bt}. Even at core bounce and immediately
afterward, the density profile is so shallow that the ordinary H- and
L-resonances may occur within the collective neutrino
region~\cite{Duan:2007sh}. In this case the effects discussed here are
irrelevant because the mu-tau matter effect is negligible.  Probably
our effects are also negligible during the cooling phase of an
iron-core SN. However, flavor oscillation effects are probably largest
during the accretion phase of an iron-core SN where the flavor
dependence of the spectra and fluxes is more pronounced than during
the cooling phase~\cite{Keil:2002in}. 

When it is important, the mu-tau matter effect adds one more layer of
complication to the already vexed problem of collective SN neutrino
oscillations. It was previously recognized that ``ordinary''
collective oscillations are almost completely insensitive to the
smallness of $\Theta_{13}$ as long as it is not exactly zero. Here we
have found the opposite for the large mixing angle $\Theta_{23}$ that
is often assumed to be maximal. Even small deviations from maximal
23-mixing can imprint themselves in the collective oscillation
effect. Both results are counter-intuitive and opposite to ordinary
flavor oscillations.


\begin{acknowledgments}
We thank E.~Akhmedov for an illuminating correspondence,
  A.~Mirizzi for comments on the manuscript, and B.~Dasgupta and
  A.~Dighe for helpful comments and for making their
  manuscript~\cite{Dasgupta07} available before completion.  This
work was partly supported by the Deutsche Forschungsgemeinschaft
(grant TR-27 ``Neutrinos and Beyond''), by the Cluster of Excellence
``Origin and Structure of the Universe'' (Garching and Munich), by the
European Union (contracts No.\ RII3-CT-2004-506222 and
MRTN-CT-2004-503369), and by the Spanish grants FPA2005-01269 (MEC)
and ACOMP07-270 (Generalitat Valenciana).  AE~was supported by an FPU
grant from the Spanish Government. SP~and RT were supported by MEC
contracts ({\em Ram\'{o}n y Cajal} and {\em Juan de la Cierva},
respectively).
\end{acknowledgments}


\raggedright

\end{document}